\documentstyle[12pt,psfig]{article}
\begin{document}

\newcommand{\lsim}{\stackrel{\textstyle <}{_\sim}}
\newcommand{\okgr}{\Omega_{\rm K}}
\newcommand{\gsim}{\stackrel{\textstyle >}{_\sim}}
\newcommand{\mevt}{{\rm MeV/fm}^3}
\newcounter{sctn}
\newcounter{subsctn}[sctn]
\newcommand{\sctn}[1]{~\\ \refstepcounter{sctn} {\bf \thesctn~~ #1} \\ }
\newcommand{\subsctn}[1]
{~\\ \refstepcounter{subsctn} {\bf \thesctn.\thesubsctn~~ #1}\\}

\newcommand{\lbl}{\begin{flushright} LBNL--40217 \\[5ex] \end{flushright}}
\newcommand{\tit}{\bf Signal for the Quark-Hadron Phase Transition in 
Rotating Hybrid Stars}
\newcommand{\doe}
{This work was supported by the Director, Office of Energy Research,
Office of High Energy and Nuclear Physics, Division of Nuclear Physics,
of the U.S. Department of Energy under Contract DE-AC03-76SF00098.}

\newcommand{\autha} {F. Weber}
\newcommand{\authb} {N. K. Glendenning}
\newcommand{\authc} {S. Pei}

\newcommand{\adrb} {{\em Nuclear Science Division and\\ Institute for
    Nuclear \& Particle Astrophysics\\ Lawrence Berkeley National
    Laboratory\\ Berkeley, California 94720}\\[2ex] }

\newcommand{\adrc} {{\em Beijing Normal University\\ Department of
    Physics\\ Beijing 199875\\ P. R. China}\\[6ex]}

\newcommand{\adra} {{\em Ludwig-Maximilians University of Munich
    \\Institute for Theoretical Physics\\ Theresienstr. 37\\ 80333
    Munich \\ Germany}\\[2ex]}

\begin{titlepage}
\lbl
\begin{center}
\begin{Large}
\renewcommand{\thefootnote}{\fnsymbol{footnote}}
\setcounter{footnote}{1}
\tit \\[7ex]
\end{Large}
\renewcommand{\thefootnote}{\fnsymbol{footnote}}
\setcounter{footnote}{2}
\begin{large}
\autha~\\
\end{large}
\adra
\begin{large}
\authb~\\
\end{large}
\adrb
\begin{large}
\authc~\\
\end{large}
\adrc
\end{center}

\begin{abstract}
  For the past 20 years it had been thought that the coexistence phase
  of the confined hadronic and quark matter phases, assumed to be a
  first order transition, was strictly excluded from neutron stars.
  This, however, was due to a seemingly innocuous idealization which
  has approximated away important physics.  The reason is that neutron
  stars constitute multi-component bodies rather than single-component
  ones formerly (and incorrectly) used to describe the deconfinement
  phase transition in neutron stars.  So, contrary to earlier claims,
  `neutron' stars may very well contain quark matter in their cores
  surrounded by a mixed-phase region of quark and hadronic matter.
  Such objects are called {\sl hybrid} stars. The structure of such
  stars as well as an observable signature that could signal the
  existence of quark matter in their cores are discussed in this
  paper.
\end{abstract}
\end{titlepage}

\renewcommand{\thefootnote}{\arabic{footnote}}
\setcounter{footnote}{0}


\section{Introduction}\label{sec:intro}

\begin{figure}[h]
\begin{center}
\leavevmode
\psfig{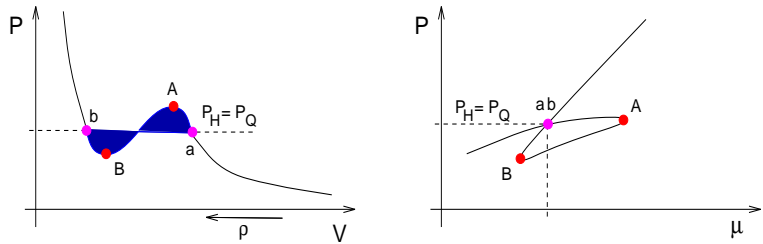}
\caption[Phase transition in a simple body]{Phase transition in a 
  `simple' body (only one conserved entity) for which pressure stays
  constant in the quark--hadron transition region `a--b'. Left: volume
  dependence of pressure for a given temperature. Points `A' and `B'
  denote metastable states. Right: same as left-hand side, but for the
  chemical potential as independent variable. The labels refer to the
  same points as in the figure to the left. Phase equilibrium `a--b'
  is mapped onto the single point `a\,b' where the curves intersect.}
\label{fig:Phasetr1}
\end{center}
\end{figure}

In all earlier work of the last two decades on the quark--hadron phase
transition in neutron stars, a degree of freedom was frozen out which
yielded a description of the transition as a {\sl constant} pressure
one, as illustrated schematically in Fig.~\ref{fig:Phasetr1}.  Since
pressure in a star is monotonically decreasing (because of hydrostatic
equilibrium), this had the explicit consequence of excluding the
coexistence phase of hadrons and quarks (region a--b in
Fig.~\ref{fig:Phasetr1}) from neutron stars. The degree of freedom
that was frozen out is the possibility of reaching the lowest possible
energy state by rearranging electric charge between the regions of
hadronic matter and quark matter in phase equilibrium.  Because of
this freedom the pressure in the mixed phase {\sl varies} as the
proportions of the phases, and therefore the coexistence phase is not
excluded from the star \cite{glen91:pt,glen97:book,hermann96:a}.

\begin{figure}[tb]
\begin{center}
\leavevmode
\psfig{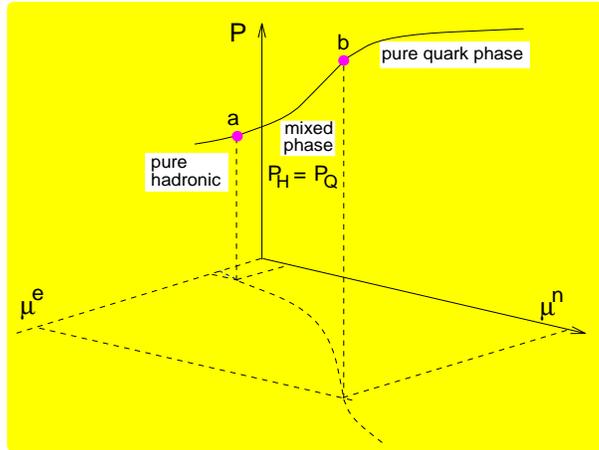}
\caption[Phase transition in a body with two conserved charges]{Phase 
  transition in a body (in our case a neutron star) with {\sl two} conserved
  entities, that is, electric charge and baryon charge. In contrast to the
  phase transition in a body with only one conserved charge, shown in
  Fig.~\protect{\ref{fig:Phasetr1}}, here pressure in the mixed phase varies
  with density. Therefore the mixed phase is not excluded from neutron stars,
  as is the case for the equation of state shown in
  Fig.~\protect{\ref{fig:Phasetr1}}.}
\label{fig:Phasetr2}
\end{center}
\end{figure}

The physical reason behind this is the conservation of baryon charge
and electric charge in neutron star matter.  Correspondingly, there
are two chemical potentials -- one associated with baryon charge and
the other associated with electric charge -- and therefore the
transition of baryon matter to quark matter is to be determined in
three-space spanned by pressure and the
 chemical potentials of the electrons and neutrons (rather
than two-space), as schematically illustrated in Fig.~\ref{fig:Phasetr2}.
 This circumstance has not been realized in the numerous
investigations published on this topic earlier \cite{glen91:pt}.

\goodbreak
\section{Three neutron star models}\label{sec:models}

To explore the implications of the mixed phase for the structure of neutron
stars, we shall employ a collection of different models for the equation of
state derived for three different assumptions about the composition of
`neutron' star matter.

\goodbreak
\subsection{Neutron stars}

In the most primitive conception, a neutron star is constituted from
neutrons. At a slightly more accurate representation, a beta stable
compact star will contain neutrons and a small number of protons whose
charge is balanced by leptons.  We represent the interactions among
baryons in the relativistic mean field theory. Details can be found
elsewhere \cite{glen85:b,weber89:e,weber91:d}. The coupling constants
in the theory are chosen so that for symmetric nuclear matter, the
five important bulk properties (energy per baryon, incompressibility,
effective nucleon mass, asymmetry energy, saturation density) are
reproduced \cite{glen86:b,schaab95:a}.

\goodbreak
\subsection{Hyperon stars} 

At the densities in the interior of neutron stars, the neutron chemical
potential will exceed the mass (modified by interactions) of various members of
the baryon octet \cite{glen82:a}. So in addition to neutrons, protons and
electrons, neutron stars are expected to have populations of hyperons which
together with nucleons and leptons are in a charge neutral equilibrium state.
Interactions among the baryons are incorporated, and coupling constants chosen,
as above \cite{glen85:b,weber89:e,weber91:d,glen86:b}.  Hyperon coupling
constants are chosen: (1) to reproduce the binding of the lambda in nuclear
matter, (2) to be compatible with hypernuclei, and (3) to support at a minimum
a neutron star of at least $1.5\,M_\odot$ \cite{glen91:pt}.

\goodbreak
\subsection{Hybrid stars} 

\begin{figure}[tb]
\begin{center}
  \leavevmode \psfig{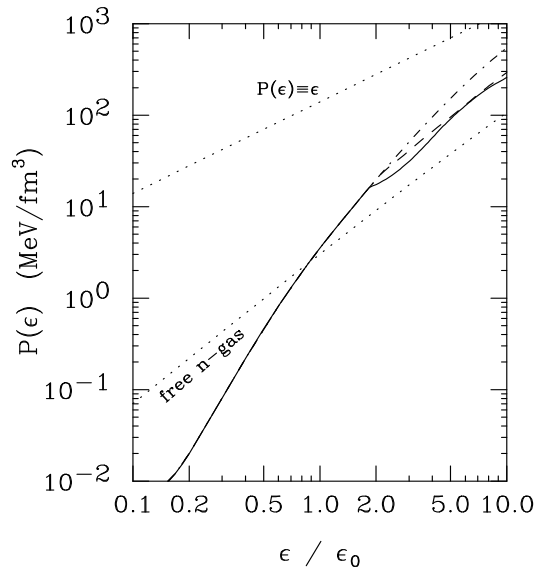}
  \caption[Graphical illustration of equations of state]{Three models for the
    equation of state. Solid curve: equation of state of hybrid star ($G_{\rm
      B180}^{\rm K240}$), dashed curve: hyperon star ($G_{\rm M78}^{\rm
      K240}$), dash-dotted curve: neutron star ($G_{\rm M78}^{\rm K240}$),
    protons and neutrons only). (For details, see
    Ref.~\protect{\cite{schaab95:a}}.)}
\label{fig:eos}
\end{center}
\end{figure}

How to handle phase equilibrium in dense neutron star matter having
two conserved charges, baryon and electric, is described in detail
elsewhere \cite{glen91:pt}. As we have seen qualitatively in
Section~\ref{sec:intro}, the properties of a phase transition in a
multi-component system is very different from the familiar one of a
single-component system formerly (and incorrectly) used to describe
the deconfinement phase transition in neutron stars
\cite{glen91:pt,glen97:book}. Models for the equation of state of neutron,
hyperon, and hybrid star matter are shown in Fig.~\ref{fig:eos}.
 One sees that the transition of confined baryonic matter to quark
matter sets in at about twice nuclear matter density $\epsilon_0$
$(=140~\mevt)$, which leads to an additional softening of the equation of
state. Pure quark matter is obtained for densities $\gsim 7 \epsilon_0$.

\goodbreak
\section{Sequences of rotating stars with constant \hfill\break baryon number}
\label{sec:seq}

Neutron stars are objects of highly compressed matter so that the
geometry of space-time is changed considerably from flat space-time.
Thus for the construction of realistic models of rapidly rotating
pulsars one has to resort to Einstein's theory of general relativity.
In the case of a star rotating at its {\sl absolute} limiting
rotational period, that is, the Kepler (or mass-shedding) frequency,
Einstein's field equations, 
\begin{equation}
  {\cal R}^{\kappa\lambda} - {1\over 2} \, g^{\kappa\lambda} \, {\cal
    R} = 8\, \pi \, {\cal T}^{\kappa\lambda}(\epsilon, P(\epsilon)) \,
  ,
 \label{eq:einstein}
\end{equation}
are to be solved selfconsistently in combination with the 
general relativistic expression which describes the onset of mass-shedding 
at the equator \cite{weber91:d,friedman86:a,glen93:drag}:
\begin{equation}
  \okgr =  \omega +\frac{\omega^\prime}{2\psi^\prime} +e^{\nu
    -\psi} \sqrt{ \frac{\nu^\prime}{\psi^\prime} +
    \Bigl(\frac{\omega^\prime}{2 \psi^\prime}e^{\psi-\nu}\Bigr)^2 } \, .
  \label{eq:okgr}
\end{equation} The metric for a rotating star, suitable for both the
interior and exterior, reads
\cite{friedman86:a,butterworth76:a,datta88:a},
\begin{eqnarray}
  ds^2 = - \,e^{2\nu} dt^2 + e^{2\lambda} dr^2 + e^{2\mu} d\theta^2 +
  e^{2\psi}[d\phi - \omega\,dt]^2 \; .
\label{eq:metric}
\end{eqnarray} Because of the underlying symmetries, the metric
functions $\nu$, $\psi$, $\mu$, and $\lambda$ are independent of $t$
and $\phi$ but depend on $r$, $\theta$ and $\Omega$.  The $\omega$
denotes the angular velocity of the local inertial frames (frame
dragging frequency) and depends on the same variables as the metric.
The frequency $\bar\omega \equiv \Omega-\omega(r,\theta,\Omega)$, is
the star's rotational frequency relative to the frequency of the local
inertial frames, and is the one on which the centrifugal force acting
on the mass elements of the rotating star's fluid depends
\cite{hartle67:a}.  The quantities ${\cal R}^{\kappa\lambda}$,
$g^{\kappa\lambda}$, and ${\cal R}$ denote respectively the Ricci
tensor, metric tensor, and Ricci scalar (scalar curvature). The
dependence of the energy-momentum tensor ${\cal T}^{\kappa\lambda}$ on
pressure and energy density, $P$ and $\epsilon$ respectively, is
indicated in Eq.~(\ref{eq:einstein}).  The primes in (\ref{eq:okgr})
denote derivatives with respect to Schwarzschild radial coordinate,
and all functions on the right are evaluated at the star's equator.
All the quantities on the right-hand side of (\ref{eq:okgr}) depend
also on $\okgr$, so that it is not an equation for $\okgr$, but a
transcendental relationship which the solution of the equations of
stellar structure, resulting from Eq.~(\ref{eq:einstein}), must
satisfy if the star is rotating at its Kepler frequency.  (Details can
be found in \cite{weber91:d}.)

The outcome of two selfconsistent calculations, one for a hybrid and
the other for a conventional hyperon star, is compared in
Figs.~\ref{fig:freq.hybrid} and \ref{fig:freq.ns}. The stars' baryon
number is kept constant during spin-down from the Kepler frequency to
zero rotation, as it should be.
\begin{figure}[ht] 
\parbox[t]{6.5cm}
{\leavevmode
  \psfig{figure=freq.hybrid.ps.bb,width=6.95cm,height=7.2cm,angle=90}
  {\caption[Frequency dependence of hyperon thresholds]{Frequency dependence of
      quark structure in rotating hybrid stars.  The radial direction is along
      the star's equator (solid curves) and pole (dashed curves). The
      nonrotating star mass is $\sim 1.42\,M_\odot$.  The rotational frequency
      ranges from zero to Kepler.}
\label{fig:freq.hybrid}}}
\ \hskip 1.4cm   \
\parbox[t]{6.5cm}
{\leavevmode
  \psfig{figure=OkEq_1.40_HV.book.ps.bb,width=6.45cm,height=7.2cm,angle=0}
  {\caption[Frequency dependence of hyperon threshold]{Same as
      Fig.~\protect{\ref{fig:freq.hybrid}}, but for a conventional hyperon star
      ($M\sim 1.40\,M_\odot$), that is, transition to quark matter is
      suppressed. The $\Sigma^-$ is absent for $\Omega\gsim 4700~{\rm s}^{-1}$
      because the central density falls below the threshold density of the
      $\Sigma^-$ particle.}
\label{fig:freq.ns}}}
\end{figure} The frequency $\Omega$ is assumed to be constant
throughout the star's fluid since uniform rotation is the
configuration that minimizes the mass-energy at specified baryon
number and angular momentum \cite{hartle67:b}.

According to the mass of a hybrid star, it may consist of an inner
sphere of purely quark matter (lower-left portion of
Fig.~\ref{fig:freq.hybrid} for which $\Omega\lsim 1250~{\rm s}^{-1}$
and $r\lsim 4.5$~km), surrounded by a few kilometers thick shell of
mixed phase of hadronic and quark matter arranged in a lattice
structure, and this surrounded by a thin shell of hadronic liquid,
itself with a thin crust of heavy ions \cite{glen97:book}.  The
Coulomb lattice structure of varying geometry introduced to the
interior of neutron stars \cite{glen97:book}, which may have dramatic
effects on pulsar observables including transport properties and the
theory of glitches, is a consequence of the competition of the Coulomb
and surface energies of the hadronic and quark matter phase. This
competition establishes the shapes, sizes and spacings of the rarer
phase in the background of the other (that is, for decreasing density:
hadronic drops, hadronic rods, hadronic plates immersed in quark
matter followed by quark plates, quark rods and quark drops immersed
in hadronic matter) so as to minimize the lattice energy.  For an
investigation of the structure of the mixed phase of baryons and
quarks predicted by Glendenning, we refer to \cite{glen95:a}.

As a rotating hybrid star spins down it becomes less deformed and the
central density {\sl rises}. For some pulsars the mass and initial
rotational frequency $\Omega$ may be such that the central density
rises from below the critical density for dissolution of baryons into
their quark constituents. This is accompanied by a sudden shrinkage of
the hybrid star, which dramatically effects its moment of inertia and
hence the braking index of a pulsar, as we shall see in the next
sections.

\goodbreak
\section{Moment of inertia}\label{sec:moi}

Elsewhere we have obtained an expression for the moment of inertia of
a relativistic star that accounts for the centrifugal flattening of
the star \cite{glen93:drag,glen92:crust}. It is given by
\begin{equation}
  I(\Omega) = 4\pi \int_0^{\pi/2}\! d\theta \int_0^{R(\theta)}\! d r
  \; { e^{\lambda+\mu+\nu+\psi} \; {[\epsilon + p]} \over {e^{2(\nu -
        \psi)} - \bar\omega^2} } \; { {\Omega - \omega}\over \Omega }
  \, .
\label{eq:idef}
\end{equation} The radial distribution of energy density and pressure,
$\epsilon(r) {\rm~and~} p(r)$, are found from the solution of the
equations for rotating relativistic stars.  For slow rotation this
expression reduces to the well known, and frequency {\sl independent}
result \cite{glen92:crust}.

\begin{figure}[tb]
\begin{center}
\leavevmode
 \psfig{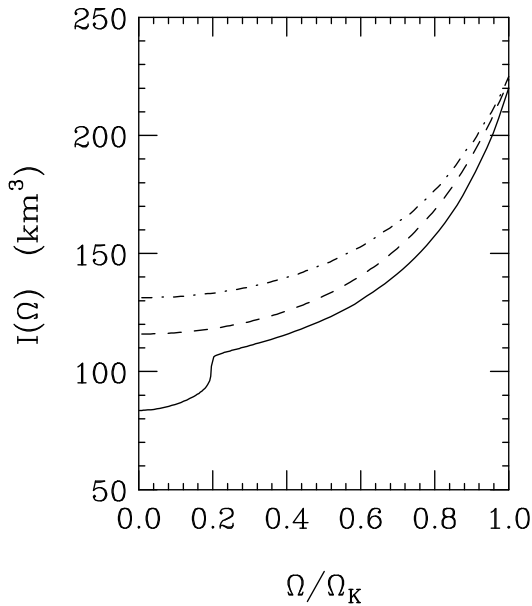}
\caption[Moment of inertia]{Moment of inertia as a function of rotational
  frequency (in units of Kepler) for three neutron stars (solid curve:
  hybrid star, dashed curve: hyperon star, dash-dotted curve: neutron
  star made up of only protons and neutrons) with different
  constitutions as described in Sect.~\protect{\ref{sec:models}}. The
  baryon number is constant along each curve. The development of a
  quark matter core for decreasing frequency (increasing density)
  causes a sudden reduction of $I$.}
\label{fig:moi}
\end{center}
\end{figure}

We show how the moment of inertia changes with frequency in
Fig.~\ref{fig:moi} for stars having the same baryon number but 
different constitutions, as described in Sect.~\ref{sec:models}. 
 The two curves without a drop at low frequencies are for
conventional neutron (dot-dashed line) and hyperon (dashed) stars of
non-rotating mass $M\sim 1.45\,M_\odot$.  The solid line is for a hybrid
star of roughly the same mass and baryon number.  The shrinkage of the
hybrid star due to the development of a quark matter core at low
frequencies, known from Fig.~\protect{\ref{fig:freq.hybrid}},
manifests itself in a sudden reduction of $I$, which is the more
pronounced the bigger the quark matter core (i.e., the smaller
$\Omega$) in the center of the star.

\goodbreak
\section{Evolution of braking index of pulsars}\label{sec:braking}

Pulsars are identified by their periodic signal believed to be due to
a strong magnetic field fixed in the star and oriented at an angle
from the rotation axis.  The period of the signal is therefore that of
the rotation of the star.  The angular velocity of rotation decreases
slowly but measurably over time, and usually the first and
occasionally second time derivative can also be measured. Various
energy loss mechanisms could be at play such as the dipole radiation,
part of which is detected on each revolution, as well as other losses
such as ejection of charged particles \cite{ruderman87:a}.  The
measured frequency and its time derivative have been used to estimate
the spin-down time or age of pulsars.  The age is very useful for
classifying and understanding pulsar phenomena such as glitch
activity.

Let us assume, as usual, that the pulsar slow-down is governed by a
single mechanism or several mechanisms having the same power law.  Let
us write the energy balance equation as
\begin{equation}
  \frac{dE}{dt} = \frac{d}{dt} \, \Bigl\{ \frac{1}{2}\, I \, \Omega^2
  \Bigr\} = - \, C \, \Omega^{n+1} \, ,
 \label{eq:engloss}
\end{equation} where, for magnetic dipole radiation, $ C= \frac{2}{3}
m^2 \sin^2 \alpha $, $n=3$ , $m$ is the magnetic dipole moment and
$\alpha$ is the angle of inclination between magnetic moment and
rotation axis. If, as is customary, the angular velocity $\Omega$ is
regarded as the only time-dependent quantity, one obtains the usual
formula for the rate of change of pulsar frequency,
\begin{equation}
  \dot{\Omega} = - \, K \, \Omega^n \, ,
\label{eq:braking}
\end{equation} with $K$ a constant and $n$, the {\sl braking index}.
From the braking law (\ref{eq:braking}) one usually defines from its
solution, the spin-down age of the pulsar
\begin{equation}
  \tau = - \, \frac{1}{n-1} \, \frac{\Omega}{\dot{\Omega}} \, .
\end{equation}

However the moment of inertia is {\sl not} constant in time but responds to
changes in rotational frequency, as can be seen in Fig.~\ref{fig:moi}, more or
less according to the softness or stiffness of the equation of state (that is,
the star's internal constitution) and according as the stellar mass is small or
large.  This response changes the value of the braking index in a frequency
dependent manner, even if the sole energy-loss mechanism were {\sl pure} dipole
as in Eq.~(\ref{eq:engloss}). Thus during any epoch of observation, the braking
index will be measured to be different from $n=3$ by a certain amount. How much
less depends, for any given pulsar, on its rotational frequency and for
different pulsars of the same frequency, on their mass and on their internal
constitution.

When the frequency response of the moment of inertia is taken into
account, Eq.~(\ref{eq:braking}) is replaced by
\begin{equation}
  \dot{\Omega}= - \, 2 I K \, \frac{\Omega^n}{ 2 I + I^{\prime}\Omega
    } = - \, K \, \Omega^n \biggl\{ 1-\frac{I^{\prime}}{2I} \, \Omega +
  \Bigl(\frac{I^{\prime}}{2I} \, \Omega \Bigr)^2 - \cdots \biggr\} \, ,
\label{eq:braking2}
\end{equation} where $I^\prime \equiv dI/d\Omega$ and $K=C/I$.  This
explicitly shows that the frequency dependence of $\dot{\Omega}$
\begin{figure}[htb]
\begin{center}
\leavevmode
\psfig{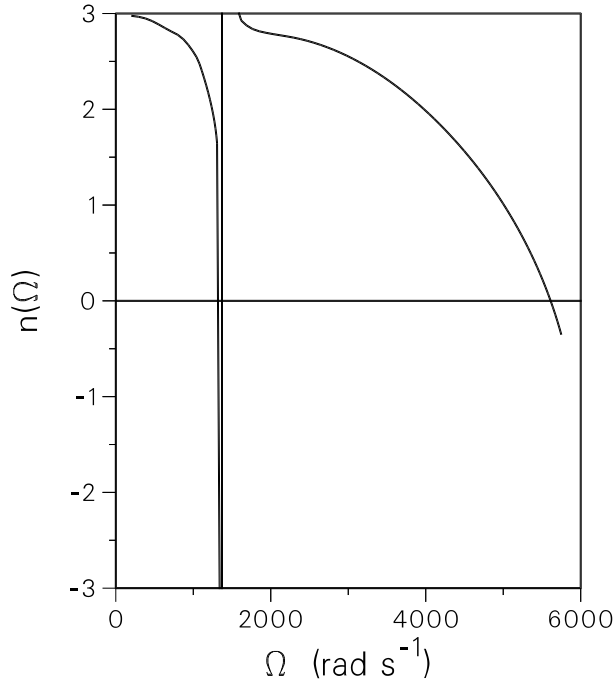}
\caption[Braking index]{Braking index as a function of rotational 
  frequency for a hybrid star. The dip at low frequencies is driven by
  the phase transition of baryonic matter into quark matter. The
  overall reduction of $n$ below 3 is due to the frequency dependence
  of $I$ and therefore holds for all three types of stars (neutron,
  hyperon, and hybrid).}
\label{fig:n}
\end{center}
\end{figure} 
corresponding to {\sl any} mechanism that absorbs (or deposits)
rotational energy such as Eq.~(\ref{eq:engloss}) cannot be a power
law, as in Eq.~(\ref{eq:braking}) with $K$ a constant. It must depend on
the mass and internal constitution of the star through the response of
the moment of inertia to rotation as in Eq.~(\ref{eq:braking2}).

Equation (\ref{eq:braking2}) can be represented in the form of
Eq.~(\ref{eq:braking}) (but now with a frequency dependent prefactor)
by evaluating
\begin{equation}  n(\Omega)=\frac{\Omega\, \ddot{\Omega}
    }{\dot{\Omega}^2} = n - \frac{ 3 \, I^\prime \, \Omega + I^{\prime
      \prime} \, \Omega^2 } {2\, I + I^\prime \, \Omega} \, .
 \label{eq:index}
\end{equation} Therefore the effective braking index depends
explicitly and implicitly on $\Omega$.  The right side reduces to a
constant $n$ only if $I$ is independent of frequency. But his cannot
be, not even for slow pulsars if they contain a quark matter core.
The centrifugal force ensures the response of $I$ to $\Omega$.  As an
example, we show in Fig.~\ref{fig:n} the variation of the braking
index with frequency for the rotating hybrid star of
Fig.~\ref{fig:freq.hybrid}.  For illustration we assume dipole
radiation.  As before, the baryon number of the star is kept constant.
Because of the structure in the moment of inertia, driven by the phase
transition into the deconfined quark matter phase, the braking index
deviates dramatically from 3 at small rotation frequencies.  Such an
anomaly in $n(\Omega)$ is not obtained for conventional neutron or
hyperon stars because their moments of inertia increase smoothly with
$\Omega$ (cf.\ Fig.~\ref{fig:moi}). The observation of such an anomaly
in the timing structure of pulsars may thus be interpreted as a signal
for the development of quark-matter cores in the centers of pulsars.
As shown in \cite{glen97:a}, the duration over which the braking index
would be anomalous may be 1/100'th of the active pulsar lifetime.
Given that $\sim 10^3$ pulsars are known (actually $> 700$ as of this
date), about 10 of these may be signaling the phase transition!

\section{Summary}

Whether or not the cores of neutron stars are in the deconfined quark
matter phase makes little difference to their static properties such
as the range of possible masses, sizes, or even their limiting
rotational frequencies. However we find that dramatic effects may
occur in the timing structure of a pulsar's spin--down, that is, in
the so-called braking index of a pulsar: the possible phase transition
of confined baryonic matter into its deconfined phase may register
itself in a dramatic change of the braking index, which is completely
absent for stars entirely made up of confined baryonic matter.  We
estimate that about 10 out of the presently known $\sim 700$ pulsars
could be signaling the phase transition.

\vskip 1.truecm
{\bf Acknowledgements}: \doe.


\clearpage

\end{document}